\documentclass[preprint,notoc]{JHEP3}
\usepackage{epsfig}
\def\eslt{\not\!\!{E_T}}
\def\to{\rightarrow}

\def\bi{\begin{itemize}}
\def\ei{\end{itemize}}

\def\tb{\tilde b}

\def\tst{\tilde t}
\def\ttau{\tilde \tau}

\def\tg{\tilde g}

\def\tq{\tilde q}
\def\tw{\widetilde W}
\def\tz{\widetilde Z}
\def\alt{\stackrel{<}{\sim}}
\def\agt{\stackrel{>}{\sim}}

\title{
Updated reach of CERN LHC  and   constraints from
relic density, $b\to s\gamma$ and $a_\mu$ in the mSUGRA model
}

\author{Howard Baer, Csaba Bal\'azs, 
Alexander Belyaev\footnote{On leave of absence from 
Nuclear Physics Institute, Moscow State University.}, 
Tadas Krupovnickas
\\ Department of Physics, Florida State University\\ 
Tallahassee, FL 32306, USA\\
E-mail: \email{baer@hep.fsu.edu},\email{balazs@hep.fsu.edu},
\email{belyaev@hep.fsu.edu}, \email{tadas@hep.fsu.edu}}
\author{Xerxes Tata
\\ Department of Physics and Astronomy, University of Hawaii,\\
Honolulu, HI 96822, USA \\ 
E-mail: \email{tata@phys.hawaii.edu}}

\preprint{\vbox{\hbox{FSU-HEP-030416} \vspace{0.2cm}
                \hbox{UH-511-1023-03}}} 

\abstract{We present an updated assessment of the reach of the CERN LHC
$pp$ collider for supersymmetric matter in the context of the minimal
supergravity (mSUGRA) model. In addition to previously examined
channels, we also include signals with an isolated photon or with a
leptonically decaying $Z$ boson. 
For an integrated luminosity of $100$ fb$^{-1}$,
values of $m_{1/2}\sim 1400$ GeV can be probed for small $m_0$,
corresponding to a gluino mass of $m_{\tg}\sim 3$ TeV.  For large $m_0$,
in the hyperbolic branch/focus point region, $m_{1/2}\sim 700$ GeV can
be probed, corresponding to $m_{\tg}\sim 1800$ GeV.  We also map out
parameter space regions preferred by the measured values of the dark
matter relic density, the $b\to s\gamma$ decay rate, and the muon
anomalous magnetic moment $a_\mu$, and discuss how SUSY might reveal
itself in these regions.
We find the CERN LHC can probe the entire stau co-annihilation
region and also most of the heavy Higgs annihilation funnel allowed by WMAP
data, except for some range of large $m_0$ and $m_{1/2}$ if
$\tan\beta \agt 50$.
}


\begin{document}

\section{Introduction}
\label{sec:intro}

The search for supersymmetric (SUSY) matter is one of the primary
objectives of experiments at high energy colliders\cite{reviews}.  SUSY
matter may reveal itself through indirect effects\cite{indirect}, as in
contributions to rare decays such as $b\to s\gamma$, $b \to
s\ell\bar{\ell}$ or $B_s\to\mu^+\mu^-$, or via contributions to electric
or magnetic moments such as the dipole electric moment of the electron
or neutron or $(g-2)_\mu$. It is possible that relic SUSY
cold dark matter (CDM) has already been detected gravitationally, and
recent analyses of WMAP and other data sets indicate that the relic
density $\Omega_{CDM}h^2
=0.1126^{+0.0161}_{-0.0181}$(2$\sigma$ CL)\cite{wmap}, with a
baryonic density about six times smaller.  
Both direct and indirect searches for relic SUSY CDM
are underway\cite{cdmreview}.  Detection of a signal in any of these
experiments would provide evidence for physics beyond the standard model, but
the results may have both supersymmetric as well as non-supersymmetric
interpretations.

The definitive discovery of SUSY matter will likely have to come from
experiments operating at high energy colliders. Already, negative searches 
for SUSY by the LEP2 experiments have resulted in significant bounds: for 
instance, the light chargino $\tw_1$ must have mass 
$m_{\tw_1}>103.5$ GeV\cite{lep2w1}, while a SM-like higgs boson
must have mass $m_h>114.1$ GeV\cite{lep2h}.
The reach of the Fermilab Tevatron collider has been examined as well.
In the context of the minimal supergravity model (mSUGRA)\cite{msugra},
with parameters
\begin{equation}
m_0,\ m_{1/2},\ A_0,\ \tan\beta ,\ sign(\mu ) ,
\end{equation}
$m_{1/2}$ values of up to 250~GeV (corresponding to $m_{\tg}\alt
600$~GeV) can be probed with 25~fb$^{-1}$ for $m_0 \alt
200$~GeV\cite{tevreach} and small values of $\tan\beta$.  Then, the best
reach is obtained via the clean trilepton channel.  The reach is
considerably reduced for larger values of $m_0$ or for large values of
$\tan\beta$\cite{new3l,bk,mp}.

The CERN LHC is expected to begin operating in 2008 with $pp$
collisions at $\sqrt{s}=14$ TeV. While the initial luminosity
is expected to be $\sim 10$~fb$^{-1}$ per year, 
an integrated luminosity of several hundred
 fb$^{-1}$ is ultimately anticipated. The reach of the LHC for
supersymmetric matter has been evaluated in the mSUGRA model for
low\cite{bcpt,ac,cms,atlas} and high\cite{ltanblhc,ac,cms} 
values of $\tan\beta$, for slepton\cite{slepton} and chargino-neutralino
production{\cite{tril},
and even in the experimentally unfavorable case where the lightest
supersymmetric particle (LSP) decays hadronically
via $R$-parity violating interactions\cite{rviol}.
The LHC reach has also been evaluated for models with non-universal soft 
terms as given by non-minimal $SU(5)$ SUSY GUTs\cite{nonminreach} 
and gaugino mediation
models\cite{inoMSB}. 
Finally, the LHC reach has been evaluated for gauge-mediated\cite{gmsb} (GMSB)
and anomaly-mediated (AMSB) SUSY breaking models\cite{amsb}.

In studies evaluating the reach of the LHC, the signal channels have been
classified by the number of isolated leptons present in each event.
The isolated
leptons usually arise as end products of cascade decays of gluinos, squarks
or other massive SUSY particles\cite{decays}. At large values of 
the parameter $\tan\beta$, the $b$ and $\tau$ Yukawa couplings become
large, so that cascade decays to final states containing $b$-jets
and $\tau$-leptons are enhanced\cite{ltanb}.

In this paper, we update our reach projections for the CERN LHC for 
several reasons.
\begin{itemize}
\item We calculate the sparticle production and cascade decay
events using ISAJET v7.64\cite{isajet}. This version includes a variety
of radiative corrections and improvements to the sparticle mass
spectrum that were not present in earlier ISAJET versions,
including two-loop evolution of all RGEs. 
ISAJET 7.64 gives good agreement with spectra generated by
the Suspect, SoftSUSY and Spheno codes, as compiled by
Allanach, Kraml and Porod\cite{kraml}.
In addition, 3-body decay matrix elements are included\cite{new3l,nojiri},
so that decay product energy distributions are more accurately modeled.
\item We adopt the code CMSJET v4.801 to model the CMS detector.
This gives a more accurate portrayal of CMS than the toy detector models
used in Ref. \cite{bcpt}.
\item We include additional channels to our reach projections, including
events containing a reconstructed $Z\to \ell^+\ell^-$ candidate
($\ell=e$ or $\mu$)\cite{btw_z}, and events containing isolated photons.
The photonic events may arise from radiative neutralino
decay\cite{z2z1g,tadas2} $\tz_2\to\tz_1 \gamma$ which is enhanced in the
low $\mu$\cite{bcpt} ``hyperbolic branch'' or ``focus point''
(HB/FP)\cite{hyper,fpoint} region of parameter space.

\item Our parameter space scans extend over a wider range than earlier 
reach projections. We are motivated to do so to cover the HB/FP region
with small $\mu$ where SUSY phenomenology can be significantly
different. Further, we adopt a higher integrated luminosity value
of 100 fb$^{-1}$ than our earlier studies; this integrated
luminosity should be achieved after several years
of LHC operation. 

\item We identify regions of mSUGRA parameter space consistent
with recently updated constraints\cite{indirect}
from $b\to s\gamma$, $(g-2)_\mu$ and 
$\Omega_{\tz_1}h^2$ calculations as well as the most recent
constraints from the LEP2 experiments, and where there should be
observable signals at the LHC.
\end{itemize}

In Sec. \ref{sec:details}, we present
the details of our computer calculations, cuts and detector simulation.
In Sec. \ref{sec:results}, we show our results for the LHC reach projections 
in the $m_0\ vs.\ m_{1/2}$ plane, and identify parameter regions
consistent with all exerimental
constraints. We summarize our results in Sec.~\ref{sec:conclude}.

\section{Calculational details}
\label{sec:details}

We use the CMSJET (version 4.801) fast MC package\cite{cmsjet} for the 
CMS detector response simulation. Both SM background and signal events were 
generated using ISAJET 7.64 which has undergone numerous upgrades since
ISAJET 7.37 used in our last study\cite{ltanblhc}. The improvements
include the incorporation of matrix elements for the calculation of
three body decays of gluinos, charginos and neutralinos, two loop
renormalization group evolution of all couplings and soft SUSY breaking
parameters, inclusion of 1-loop self energies for third generation
fermions, and improved evaluation of $m_A$ which, in turn, significantly
moves the boundary of the allowed parameter space of the mSUGRA model.

We have computed SM backgrounds from the following sources : $t\bar t$, QCD 
$2\to 2$ including $c\bar{c}$ and $b\bar{b}$ production, 
$W+jets$ and $Z+jets$. Backgrounds from vector boson 
pair production are
negligible to the jetty signals that we consider in this
study\cite{bcpt}.
For the photonic signal we also included backgrounds from $W\gamma$ and
$Z\gamma$ production, but found these to be negligible after hard
cuts. Since the cross section of events with
low $p_T$ is much larger than that of events with high $p_T$, we generated 
the background events in several bins of $p_T$. We followed the division 
described in \cite{ac}.
We have also adopted the cuts suggested in this study for 
$E_T^{miss}$, 0 lepton, 1 lepton, 2 $OS$ and $SS$ lepton, 3 lepton and $\ge 4$ 
isolated lepton signals. 
We regard a lepton or a photon to be isolated if
\begin{itemize}
\item it has no charged particle with $p_T > 2$ GeV in a cone with
$\Delta R < 0.3$ around the direction of the lepton.
\item $\Sigma E_T^{cell}$ in the region with $0.05 < \Delta R < 0.3$ around 
the lepton's direction has to be less than 10\% of the lepton transverse 
energy.
\end{itemize}

For the convenience of the reader, we 
present these cuts in this paper.
All events have to pass the following pre-cuts:
\begin{itemize}
\item $E_T^{miss} > 200$ GeV;
\item Number of jets, $N_j \ge 2$.
\end{itemize}
We use a modified UA1-jetfinder routine GLOBJF, 
implemented in CMSJET, to identify
calorimeter jets. A cluster of particles is labeled as a jet if it has 
transverse momentum $p_T$ greater than 40 GeV and $|\eta|<3$.
Leptons are required to satisfy the following pre-cuts:
\begin{itemize}
\item $p_T > 10$ GeV for the muons, $p_T > 20$ GeV for the electrons, 
$|\eta| < 2.4$ for both muons and electrons.
\item Electrons have to be isolated. Muon isolation is not required as
part of the pre-cuts. We call an electron or muon 
non-isolated even if it satisfies the lepton isolation criteria
but is part of a jet, or if it is isolated in the 
calorimeter, but non-isolated in the tracker. Naturally, if it is not 
isolated in the calorimeter then the lepton is called non-isolated.
\end{itemize}
The pre-cuts for the photons are:
\begin{itemize}
\item $p_T > 20$ GeV in $|\eta| < 2.4$. 
\item photons have to be isolated.
\end{itemize}
After events pass the pre-cuts, we impose 90\% lepton detection
efficiency for each lepton.

The events which pass the pre-cuts are divided into signal types according 
to the number of leptons (or photons for the isolated $\gamma$ signal). 
In the case of the $E_T^{miss}$ signal there can 
be any number of leptons, 0 lepton signal has no leptons, 1 lepton signal 
has 1 lepton, 2 OS lepton signal has 2 opposite sign leptons, 2 SS lepton 
signal has 2 same sign leptons, 3 lepton signal has 3 leptons, $\ge 4$ lepton 
signal has more than 3 leptons, $Z \to \ell^+ \ell^-$ signal has at least 2
OS, same flavor
leptons with the invariant mass of this pair within 
the interval $(M_Z-\Delta M_Z,M_Z+\Delta M_Z)$ ($\Delta M_Z$ is varied during 
the optimization procedure). Finally, the 
isolated $\gamma$ signal has any number of 
leptons plus at least one photon 
(the cut on the number of photons is varied during the optimization 
procedure).
Since muon isolation has not been included as part of the pre-cuts, if 
we impose the muon isolation during the optimization procedure, the number 
of events for some signal types can change.

A signal in any channel is considered to be observable if after our
optimization procedure described below,
\begin{itemize}
\item the number of signal events $S \ge 10$ for an integrated
luminosity of 100~fb$^{-1}$, and 
\item  $S\ge 5\sqrt{B}$, where $B$ is the corresponding number of background events.
\end{itemize}
We optimize the signal in each channel by imposing additional cuts. 
The set of cuts that we examined for this purpose are 
listed in Table \ref{cuts}. For the optimization of 
the signal $Z \to \ell^+ \ell^-$ signal, we have an additional cut: 
the invariant mass of the 
pair of opposite sign same flavor lepton pair has to be in the interval 
$(M_Z-\Delta M_Z,M_Z+\Delta M_Z)$ and $\Delta M_Z$ is taken to be 
$3, 6, 9, ..., 30$ GeV. 
For the case of the isolated photon signal, 
in addition to the optimization using the cuts in 
Table \ref{cuts}, we also vary the number of photons which can be 
$1, 2, 3, 4, \ge 5$. 
However, for large values of $m_0$ and $m_{1/2}$, events with 
$N_\gamma^{iso} > 1$ are generally too rare to pass the requirement 
of $S > 10$.

For each mSUGRA point that we analyze, we pass the various signals
through each one of the complete set of cuts just discussed. If the
signal satisfies our observability criterion for any one of these cut
choices, we consider it to be observable.\footnote{When more than one cut
choices lead to an observable signal, we retain the choice that maximizes
the
quantity $S/\sqrt{S+B}$.}


%
\begin{table}[htb]
\caption{The set of cuts that we have examined for the optimization of
the SUSY signal. Except for the muon isolation, the numbers refer to the
lower bound on the quantity listed in the first column.}
\label{cuts}
\begin{center}
\renewcommand{\arraystretch}{1.2}
\setlength{\tabcolsep}{2mm}
\begin{tabular}{|c|c|} \hline
  Variable(s) &  Values  \\ 
 \hline \hline
 N$_{j}$        &  2, 3, 4, ..., 10                                 \\
\hline
 E$_{T}^{miss}$ &  200, 300, 400, ..., 1400 GeV                      \\
\hline
 E$_{T}^{j1}$ & 40, 150, 300, 400, 500, 600, 700, 800, 900, 1000 GeV\\
 E$_{T}^{j2}$ & 40, ~80, 200, 200, 300, 300, 400, 400, 500, ~500 GeV \\
\hline
 $\Delta\phi$ $(p_{T}^{l},E_{T}^{miss})$ & 0, 20 deg.                 \\
\hline
$ Circularity $    &  0, ~0.2                                         \\
\hline
 $\mu$ isolation  &  on, off                                           \\
\hline
\end{tabular}
\end{center}
\end{table}

\section{Results}
\label{sec:results}

\subsection{Reach of the LHC in various channels}
Since sparticle masses are largely determined by the parameters $m_0$
and $m_{1/2}$ the $m_0-M_{1/2}$ plane is a convenient arena for simultaneously
displaying the SUSY reach of future experimental facilities together
with regions already excluded by current data.
Many previous calculations for the LHC reach are, therefore, presented 
in this
plane (for various choices of other parameters) 
starting with $\tan\beta =2$; this low of a $\tan\beta$ value is now largely 
excluded by the LEP2 bound on $m_h$. We begin our presentation with 
$\tan\beta =10$, in Fig. \ref{fig:10p}. The red-shaded region on the
left
is excluded
because either electroweak symmetry is not properly broken, or
$\ttau_1$ is the LSP, while that on the right is 
excluded because radiative electroweak symmetry breaking (REWSB) does
not occur.
The magenta region in the lower left is excluded by LEP2 bounds on
$m_{\tw_1}>103.5$ GeV, and $m_h>114.1$ GeV (for a SM-like light Higgs $h$).
The maximum reach is shown by the $E_T^{miss}$ contour, where the
signal events include all isolated lepton possibilities. 
Also shown for reference are contours of $m_{\tg}=2$~TeV, and
$m_{\tq}=2$~TeV. 
The maximum reach in
$m_{1/2}$ occurs for low $m_0$, where squark masses are somewhat lighter
than gluinos, so that $\tq\tq$, $\tq\tg$ and $\tg\tg$ production
processes 
all have
large rates. Gluinos as heavy as $\sim 3$ TeV may be detectable at the LHC
if squarks are somewhat lighter.
As $m_0$ increases, squark masses increase, so gluino pair
production becomes the dominant sparticle production
mechanism. At very large values of $m_0$, the squarks (and also
sleptons) essentially
decouple, and the reach contours flatten out, since $m_{\tg}$ is roughly
constant for each value of $m_{1/2}$. The maximal reach for 100 fb$^{-1}$
of integrated luminosity is $m_{1/2}\sim 700$ GeV, corresponding to
a gluino mass value of $m_{\tg}\sim 1800$ GeV. 

\FIGURE[t]{\epsfig{file=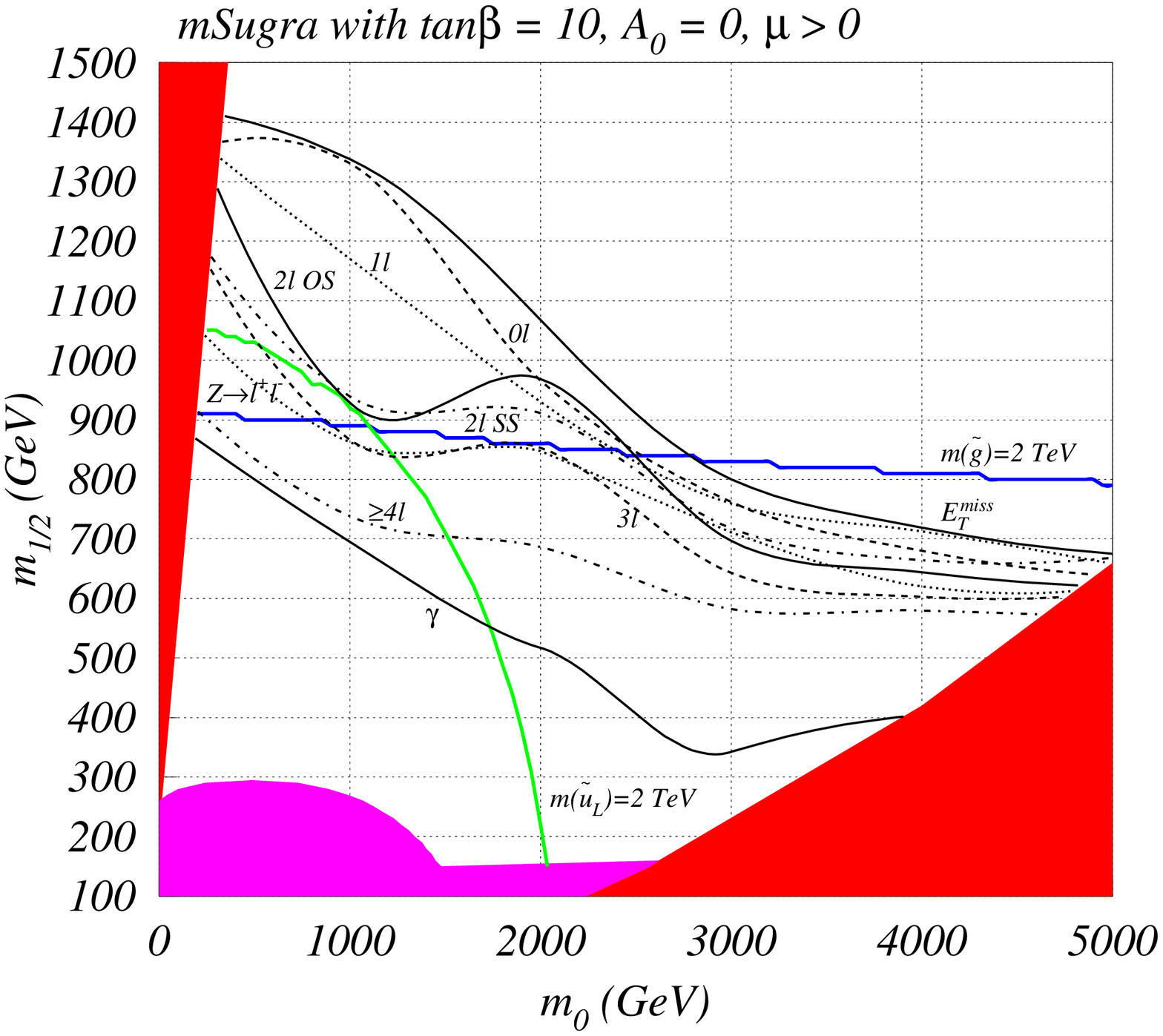,width=15cm} 
\caption{The reach of CERN LHC in the 
$m_0\ vs.\ m_{1/2}$ parameter plane 
of the mSUGRA model, with
$\tan\beta =10$, $A_0=0$ and $\mu >0$,
assuming 100 fb$^{-1}$ of integrated luminosity. The red (magenta) regions are
excluded by theoretical (experimental) constraints discussed in the
text. 
We show the reach in the $0\ell$, $1\ell$, $OS$, $SS$, $3\ell$,
$\ge 4\ell$, $\gamma$ and $Z$ channels, as well as in the
``inclusive'' $\eslt$ channel. 
}
\label{fig:10p}}

The reach for
SUSY signals in the individual channels introduced
in the last section are shown by the
various contours labelled by the corresponding topology  in Fig. \ref{fig:10p}.
For low values of $m_0$, sleptons and squarks are relatively light, 
and $\tg\tg$, $\tg\tq$ and $\tq\tq$ production processes
all occur at large rates; direct slepton and sneutrino pair production rates 
are much smaller. However, 
left-squarks $\tq_L$ decay frequently into the heavier chargino 
and heavier neutralinos, which in turn may decay to sleptons. 
These cascade decays frequently terminate in isolated leptons,
which together with leptons from decays from tops and stops in SUSY processes, 
result in large rates for leptonic signals. For a fixed value of
$m_{1/2}$, 
as $m_0$ increases, $\tg$ decay to $\tq_L$ becomes suppressed,
and $BF(\tg\to\tq_R)$ increases. Since $\tq_R\to q\tz_1$ most of the
time, 
cascade decays then give rise to a higher fraction of $\eslt +jets$
events, and the leptonic signal from the decay of gluinos is
reduced. Furthermore, since squarks become heavier
relative to gluinos as $m_0$ is increased, the leptonic signal from
directly produced $\tq_L$ also becomes smaller. 
As $m_0$ increases even further, $\tg\to \tq_R q$ also becomes suppressed
or even forbidden, and $\tg\to \tst_1 t$ (and possibly $\tg \to b\tb_{1,2}$)
dominates.
The decays through tops and stops gives rise again to leptonic states
due to $\tst_1$ and $t$ leptonic decays. This results in an increased
reach at moderate $m_0$ values via leptonic modes such as $SS$ and $OS$ 
dileptons, and $3\ell$ events.
As $m_0$ increases even more, $\tg$ two-body decays become completely
forbidden, and three-body decays dominate, and the leptonic
reach contours tend to level off. 

Finally, we show a contour that marks the signal reach for events including 
$\eslt +jets$ plus at least one isolated photon. 
In evaluating this reach, we have only retained physics backgrounds in
our calculation. Detector-dependent backgrounds where a jet fakes a
photon may be significant.\footnote{If we take the probability for a
jet to fake a photon to be $5\times 10^{-4}$ and assume that the hard
scattering events have $\sim 10$ 30-40~GeV ``jets'' in them, about one
in 500 background events will also appear to have an isolated photon. 
Assuming that this fake photon background can be estimated by
reducing the physics background in the inclusive $E_T^{miss}$ channel by
500, we find that this background is somewhat smaller, but of the same
order of magnitude as the physics background that we have evaluated.
A real evaluation of this detector-dependent background is beyond the
scope of our analysis.}
In most of the parameter 
space, the additional photon arises from $h\to\gamma\gamma$ decay, 
where the $h$ is produced copiously in sparticle cascade decays,
especially from $\tz_2\to \tz_1 h$.
In these regions, in fact, if we require $two$ isolated photons,
then we can reconstruct a di-photon invariant mass. 
This is illustrated in Fig. \ref{fig:mgg} for 
the parameter space point $m_0=m_{1/2} =500$ GeV, $A_0=0$, $\tan\beta =30$
and $\mu >0$.
It is amusing to note that the $h\to\gamma\gamma$
signal should be visible in SUSY events for 100 fb$^{-1}$ of 
integrated luminosity.  We see that (for these parameters) 
the highest possible luminosity is needed for the detection
of this severely rate-limited, but essentially SM background-free, signal. 
A detection of $h$ in this manner would nicely confirm its detection 
via $gg \to h \to \gamma\gamma$, where the signal would be picked out
as a small peak
above an enormous continuum background. A Higgs signal in the multijet
plus $\gamma\gamma$ + $E_T^{miss}$ channel would also suggest the
supersymmetric origin of the Higgs production process. 
\FIGURE[t]{\epsfig{file=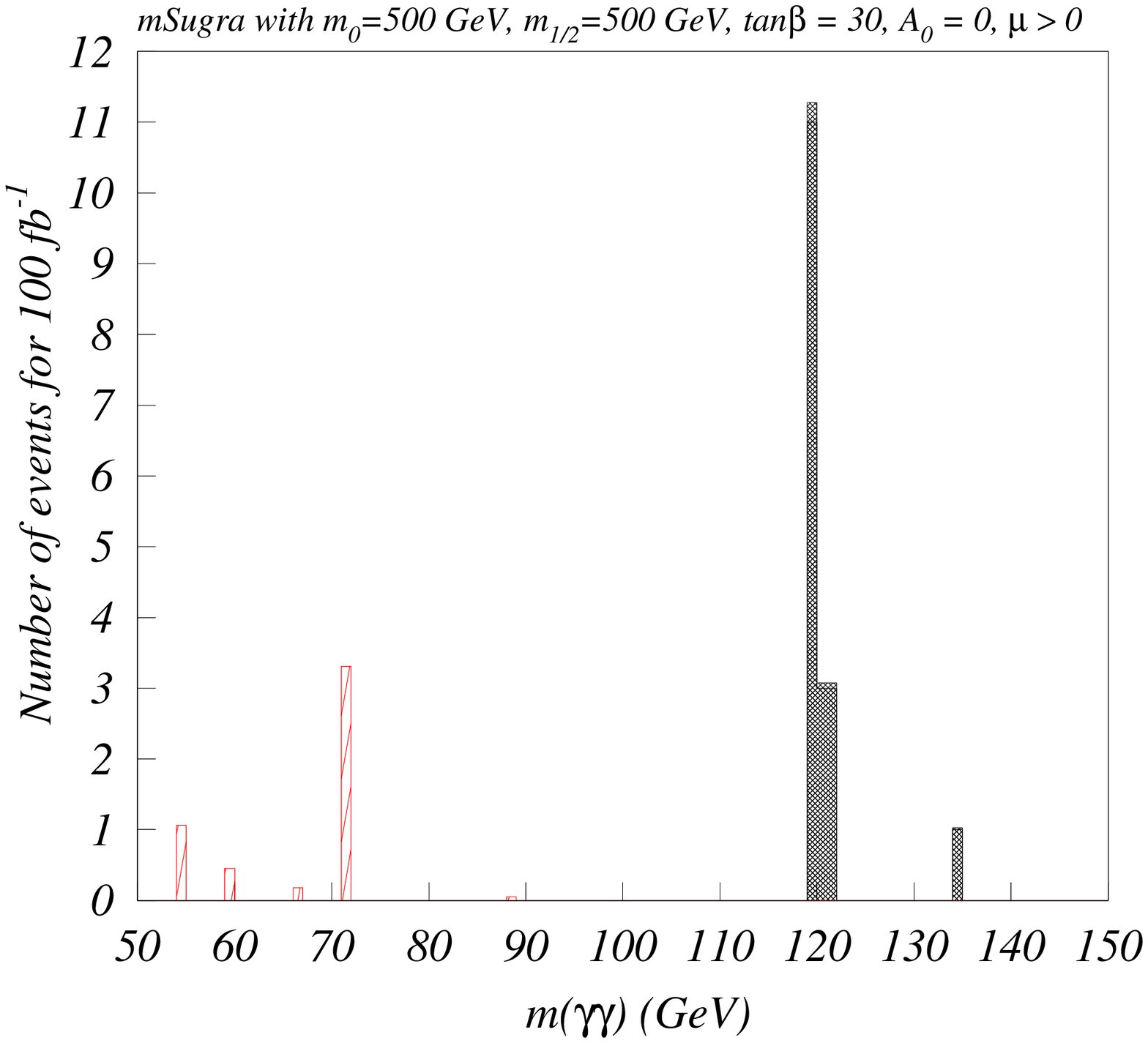,width=15cm} 
\caption{The diphoton invariant mass in SUSY events with
two isolated photons. The shaded histogram is for diphotons from SUSY
events, while the
hatched histogram represents corresponding events of SM origin. The
background has been obtained by scaling a Monte Carlo run for a lower
value of integrated lumnosity.}
\label{fig:mgg}}
As $m_0$ becomes very large, and the region of low $|\mu |$ is
approached, the radiative decay $\tz_2\to \tz_1\gamma$ becomes
enhanced\cite{tadas2}. In this HB/FP
region, the isolated photon contour likewise turns up, to reflect the
increase in isolated photon activity from neutralino radiative decay.

As we move to larger values of the parameter $\tan\beta$, as
in Fig. \ref{fig:30p} for $\tan\beta =30$, the first thing to notice is that
the left red-shaded region, where $\ttau_1$
is the LSP, has expanded. 
This is because
as $\tan\beta$ increases, both the $\tau$ (and also $b$)
Yukawa couplings become non-negligible. This results in a reduction of
stau and sbottom soft breaking masses via RGE running, 
and also in greater $L-R$ mixing, which again reduces the
mass of the lightest eigenstates.
The negative results of searches for heavy isotopes
of hydrogen (or other elements) results in limits that are many orders of 
magnitude below their expected relic density in big-bang
cosmology, so that this possibility is strongly excluded.
 Finally, the large values of $f_b$ and
$f_\tau$ together with the reduction of the corresponding
sfermion masses relative to the first two generations mentioned above, 
increases various three body sparticle decays to
$b$-quarks and $\tau$ leptons, at the expense of their
first and second generation counterparts\cite{ltanb}. 
The net effect of this is to enhance sparticle cascade 
decays to final states containing $b$-quarks and $\tau$-leptons
in the low $m_0$ region. This results in a diminution of the 
reach in isolated lepton channels at low $m_0$, as compared
with Fig. \ref{fig:10p}. The large $b$ and $\tau$ Yukawa couplings 
hardly affect the $\eslt +jets$ and $0\ell$  signals, and so the
ultimate reach of the LHC changes little in proceeding from $\tan\beta =10$
to $\tan\beta =30$. 

\FIGURE[t]{\epsfig{file=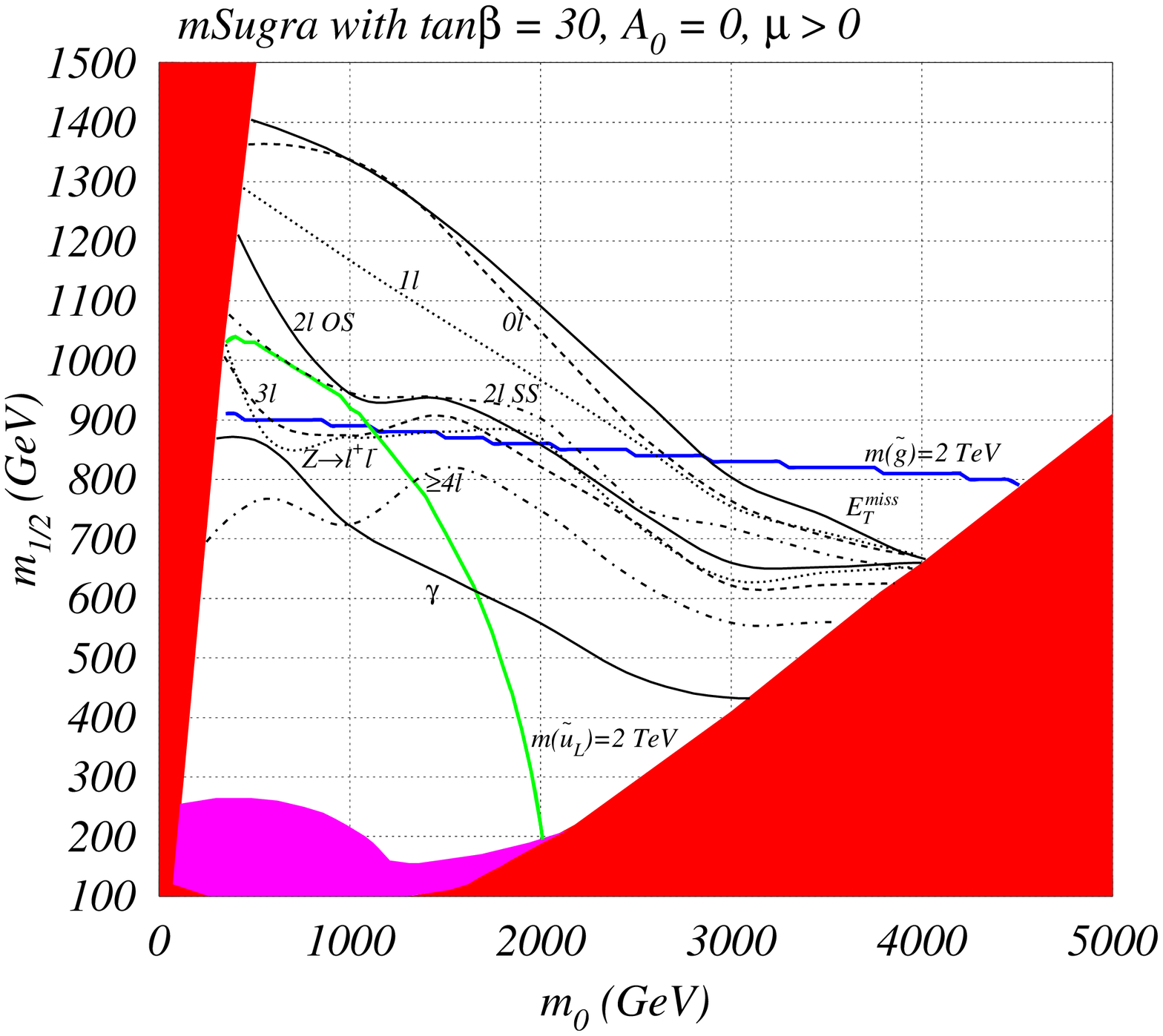,width=15cm} 
\caption{The reach of CERN LHC in the 
$m_0\ vs.\ m_{1/2}$ parameter plane 
of the mSUGRA model, with
$\tan\beta =30$, $A_0=0$ and $\mu >0$,
assuming 100 fb$^{-1}$ of integrated luminosity. The red (magenta) regions are
excluded by theoretical (experimental) constraints discussed in the
text. 
We show the reach in the $0\ell$, $1\ell$, $OS$, $SS$, $3\ell$,
$\ge 4\ell$, $\gamma$ and $Z$ channels, as well as in the
``inclusive'' $\eslt$ channel. }
\label{fig:30p}}

In Fig. \ref{fig:45m}, we show the $m_0\ vs.\ m_{1/2}$ plane for
$\tan\beta =45$ and $\mu <0$. The large $b$ and $\tau$ Yukawa coupling
effects are accentuated even more in this figure:
a larger region is excluded at low $m_0$, and the reach  
via multi-lepton channels is further diminished for small $m_0$ values
where squarks and sleptons still play a role in determining 
cascade decay patterns. However, again, the overall reach in the $\eslt +jets$
and $0\ell$ channels is hardly affected. If we increase $\tan\beta$ much
beyond about 50 for $\mu <0$, the parameter space begins to close up fast
due to a breakdown in REWSB.

\FIGURE[t]{\epsfig{file=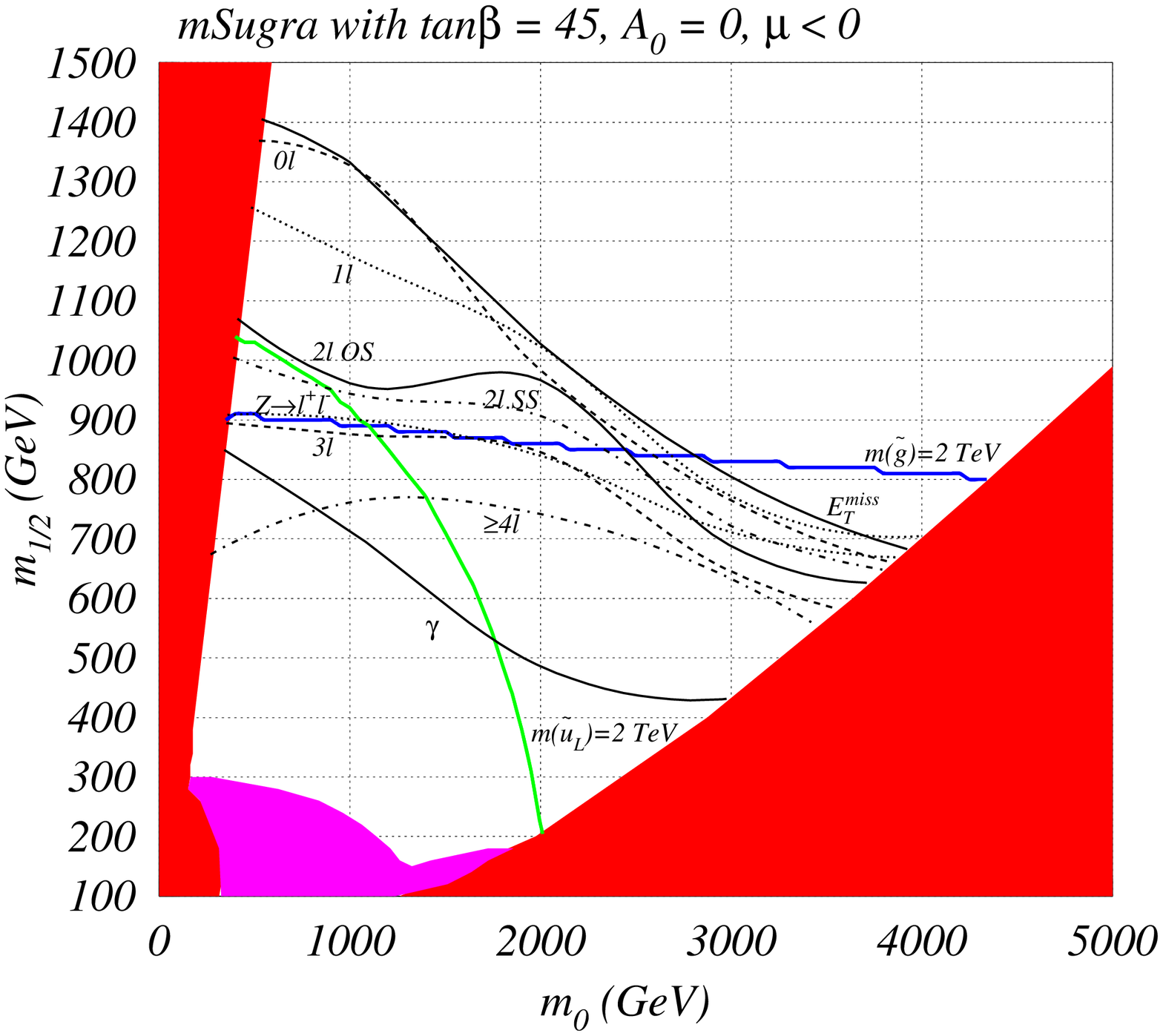,width=15cm}

\caption{The reach of CERN LHC in the 
$m_0\ vs.\ m_{1/2}$ parameter plane 
of the mSUGRA model, with
$\tan\beta =45$, $A_0=0$ and $\mu <0$,
assuming 100 fb$^{-1}$ of integrated luminosity. The red (magenta) regions are
excluded by theoretical (experimental) constraints discussed in the
text. 
We show the reach in the $0\ell$, $1\ell$, $OS$, $SS$, $3\ell$,
$\ge 4\ell$, $\gamma$ and $Z$ channels, as well as in the
``inclusive'' $\eslt$ channel. } 
\label{fig:45m}}

In Fig. \ref{fig:52p}, we show the $m_0\ vs.\ m_{1/2}$ plane for
$\tan\beta =52$ and $\mu >0$. In this case, the overall reach in the
$\eslt +jets$ and $0\ell$ channels is similar to the cases at lower
$\tan\beta$. However, in the multi-lepton channels, there is again
a suppression of reach at low $m_0$. This is because for very high
$\tan\beta$, $m_{\ttau_1}$ is so light that $\tw_1$ dominantly
decays to $\ttau_1\nu_\tau$ rather than $\tz_1 W$, and likewise,
$\tz_2\to \ttau_1\tau$ rather that $\tz_1 h$ or $\tz_1 Z$. As $m_0$
increases, the stau mass increases, and the $\tw_1\to \ttau_1\nu_\tau$
decay mode becomes more suppressed, which increases the $\tw_1\to \tz_1 W$
branching fraction. The subsequent $W$ boson decays from
$\tw_1\to \tz_1 W$ lead to hard isolated leptons.

\FIGURE[t]{\epsfig{file=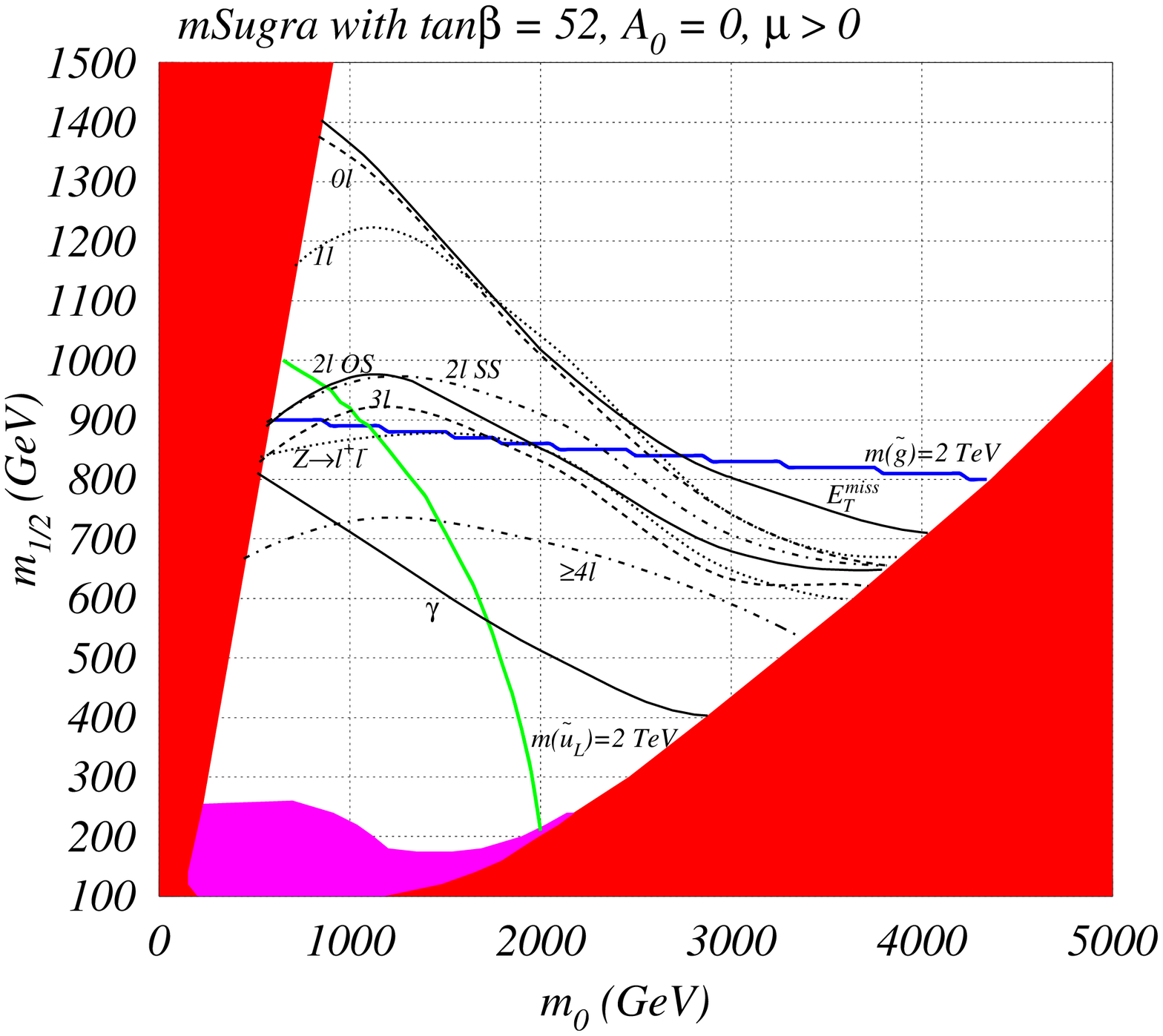,width=15cm} 
\caption{The reach of CERN LHC in the 
$m_0\ vs.\ m_{1/2}$ parameter plane 
of the mSUGRA model, with
$\tan\beta =52$, $A_0=0$ and $\mu >0$,
assuming 100 fb$^{-1}$ of integrated luminosity. The red (magenta) regions are
excluded by theoretical (experimental) constraints discussed in the
text. 
We show the reach in the $0\ell$, $1\ell$, $OS$, $SS$, $3\ell$,
$\ge 4\ell$, $\gamma$ and $Z$ channels, as well as in the
``inclusive'' $\eslt$ channel. } 

\label{fig:52p}}

Our projections of the LHC reach, where they can be directly compared,
are qualitatively similar to the results in Ref.~\cite{ac}. Differences
between the results can be attributed to the difference in the
observability criteria ($S\ge 5\sqrt{S+B}$ used in Ref.\cite{ac}
requires a minimum of 25 events to be compared with 10 events in our
study, as well as a somewhat larger significance of the signal), and to
the use
of PYTHIA instead of ISAJET for event simulation.


\subsection{The LHC reach in light of indirect constraints}

A variety of low energy measurements have been used to obtain
constraints on the parameter space of the mSUGRA model.  These include
the measured values of the cold dark matter density, the branching
fraction $BF(b\to s\gamma )$, the value of the anomalous magnetic moment
of the muon, $a_\mu =(g-2)_\mu/2$,~\footnote{There is 
considerable theoretical uncertainty in the SM value of $a_{\mu}$,
so that caution must be exercised 
in interpreting the result of experiment E821.} 
and a lower limit on
$BF(B_s\to \mu^+\mu^- )$.  Unlike collider constraints which are much
more direct, constraints from low energy measurements may be 
considerably more sensitive to details of the model, or to improvements in
the theoretical calculation.
Within a specific framework ({\it e.g.} mSUGRA), however,
these indirect constraints exclude certain regions of parameter space,
and also suggest other regions where future searches might be focussed. 
%

{\bf Neutralino relic density:} 

Measurements of galactic rotation curves, binding of galactic clusters, 
and the large scale structure of the universe all point to the need for
significant amounts of cold dark matter (CDM) in the universe. In addition, 
recent measurements of the power spectrum of the cosmic
microwave background from WMAP and other data sets\cite{wmap} lead to
\bi
\vspace{-2.6mm}
\item $\Omega_{CDM}h^2 = 0.1126^{+0.008}_{-0.009}$.
\vspace{-2.6mm}
\ei
The upper limit derived from this is a true constraint on any stable
relic from the Big Bang, such as
the lightest neutralino of the mSUGRA model. 
Regions of mSUGRA parameter space which result in a relic density that
violates this bound are excluded. A remarkable feature 
of the model is that there are
regions of the parameter space where the neutralino density lies in the
observed range, so that the neutralino makes up almost
all the cold dark matter in the universe. We remark, however, that unlike
the upper limit above,
the corresponding lower limit is flexible, since there may be additional
sources of CDM such as axions, or states associated with the hidden
sector of the mSUGRA model and/or extra dimensions.

To estimate the relic density of neutralinos in the mSUGRA model,
we use the recent calculation in Ref. \cite{bbb_relic}. 
All relevant neutralino annihilation and co-annihilation
reactions are included along with
relativistic thermal averaging\cite{gg},
which is important for obtaining the correct neutralino relic 
density in the vicinity of annihilations through $s$-channel resonances. 

{\bf BF($b\to s\gamma$):} 

The branching fraction $BF(b\to s\gamma )$ has recently been measured by
the BELLE\cite{belle}, CLEO\cite{cleo} and ALEPH\cite{aleph}
collaborations.  Combining statistical and systematic errors in
quadrature, these measurements give $(3.36\pm 0.67)\times 10^{-4}$
(BELLE), $(3.21\pm 0.51)\times 10^{-4}$ (CLEO) and $(3.11\pm 1.07)\times
10^{-4}$ (ALEPH). A weighted averaging of these results yields $BF(b\to
s\gamma )=(3.25\pm 0.37) \times 10^{-4}$. The 95\% CL range corresponds
to $\pm 2\sigma$ away from the mean. To this we should add uncertainty
in the theoretical evaluation, which within the SM dominantly comes from
the scale uncertainty, and is about 10\%.
Together, these imply the bounds,
\bi
\vspace{-2.6mm}
\item $2.16\times 10^{-4}< BF(b\to s\gamma )< 4.34 \times 10^{-4}$.
\vspace{-2.6mm}
\ei
The evaluation of the SUSY contribution to this decay entails additional
theoretical uncertainties, especially when $\tan\beta$ is large, so
that this range should be relaxed somewhat.
In our study, we show contours of $BF(b\to s\gamma )$ of
2, 3, 4 and $5\times 10^{-4}$.

The calculation of $BF(b\to s\gamma )$ used here is based upon the
program of Ref. \cite{bsg}. 
In our calculations, we also implement the running $b$-quark mass
including SUSY threshold corrections as calculated in ISAJET;
these effects can be important at large values of the 
parameter $\tan\beta$\cite{degrassi}.
Our value of the SM $b\to s\gamma$
branching fraction yields $3.4\times 10^{-4}$, with a scale uncertainty
of 10\%.

{\bf Muon anomalous magnetic moment}

The muon anomalous magnetic moment $a_\mu =(g-2)_\mu/2$ has been
recently measured to high precision by the 
E821 experiment\cite{Bennett:2002jb}:
$ a_\mu=11659204(7)(5)\times 10^{-10}$.
The most challenging parts of the SM calculation are the hadronic 
light-by-light\cite{lbl} and vacuum polarization 
(HVP)\cite{HVPee} contributions 
and their uncertainties. Presently these results are in dispute. 
In the case of 
the HVP the use of tau decay data can reduce the error, but the interpretation 
of these data is somewhat controversial\cite{HVPtau}.
Thus, the deviation of the measurement from the SM depends on which prediction 
is taken into account. According to the recent analysis by Hagiwara et 
al.\cite{HVPee}:
\bi
\vspace{-2.6mm}
\item $11.5<\delta a_\mu\times 10^{10}<60.7$. 
\vspace{-2.6mm}
\ei
A different assessment of the theoretical uncertainties\cite{HVPee}
using the procedure described in Ref.\cite{bbb} gives,
\bi
\vspace{-2.6mm}
\item $-16.7< \delta a_\mu\times 10^{10}<49.1$.
\vspace{-2.6mm}
\ei
Yet another determination has recently been made by Narison,
includes additional scalar meson loops\cite{narison}. This,
using $e^+e^-\to hadrons$ data to evaluate hadronic vacuum 
polarization contributions, yields 
\bi
\vspace{-2.6mm}
\item $-3.9<\delta a_\mu\times 10^{10}<52.1$,
\vspace{-2.6mm}
\ei
while using $\tau$-decay data results in
\bi
\vspace{-2.6mm}
\item $-26.7<\delta a_\mu\times 10^{10}<30.1$.
\vspace{-2.6mm}
\ei
The latter may include additional systematic uncertainties from
how isospin breaking effects are incorporated. 

In view of the
theoretical uncertainty, 
we only
present contours of $\delta a_\mu$, as calculated using the program
developed in \cite{bbft}, and leave it to the reader to decide the
extent of the parameter region allowed by the data. \\

{\bf $B_s\to\mu^+\mu^-$ decay}

The branching fraction of $B_s$ to a pair of muons has been
experimentally
bounded by CDF\cite{cdf}:
\bi
\vspace{-2.6mm}
\item $BF(B_s\to\mu^+\mu^- )< 2.6\times 10^{-6}$.
\vspace{-2.6mm}
\ei
If $\tan\beta \alt 20-25$ SUSY contributions to this decay are small and
do not lead to new constraints on the parameter space.
If $\tan\beta$ is large, the important SUSY
contribution to this decay is mediated by
the neutral states in the Higgs sector of supersymmetric
models. While this branching fraction is very small within the SM
($BF_{SM}(B_s \to \mu^+\mu^-)\simeq 3.4 \times 10^{-9}$), the amplitude
for the Higgs-mediated decay of $B_s$ roughly grows as $\tan^3\beta$, 
\footnote{The $\tan^3\beta$ growth obtains if the tree-level value of $m_b$
is fixed. For large values of $\tan\beta$, the radiative correction to
$m_b$ is important resulting in a deviation from the $\tan^3\beta$
growth.} 
and hence can completely dominate the SM contribution if
$\tan\beta$ is large.  
In our analysis we use the results from the last paper in Ref.~\cite{bmm} 
to delineate 
the region of mSUGRA parameters excluded by the CDF upper limit on its 
branching fraction. 

In Fig. \ref{fig:sug10p}, we again show the $m_0\ vs.\ m_{1/2}$ plot
for $\tan\beta =10$, $A_0=0$ and $\mu >0$. This time, we exhibit contours
for the low energy observables mentioned above, as mapped out
in Ref. \cite{bbb}. A $\chi^2$ analysis of the indirect constraints
was performed in Ref. \cite{sug_chi2}, which helped to identify 
regions of parameter space allowed by all the combined indirect constraints.
Several regions emerge in the $m_0\ vs.\ m_{1/2}$ plane where
the relic density can be within the WMAP range. At low $m_0$ and low
$m_{1/2}$, neutralino annihilation through $t$-channel sleptons 
can occur with a high rate. Much of this so-called ``bulk'' region is largely 
excluded by the LEP2 Higgs bound (shown as the red contour), and in 
addition, $BF(b\to s\gamma )$, though in the acceptable range, is below
its experimental central value.
In this region, sparticles are very light, and a SUSY discovery by 
the CERN LHC should be easy.

One of the remaining regions allowed by relic density constraint
is the very narrow strip adjacent to the ``$\tz_1$ not LSP'' region 
at low $m_0$,
where stau co-annihilation is important\cite{ellis}.
We see that the reach of the LHC
apparently covers much of the stau co-annihilation strip,
up to $m_{1/2}\sim 1400$ GeV for 100 fb$^{-1}$ of integrated
luminosity.
To determine the LHC reach needed to completely explore
the stau co-annihilation strip, we show the relic density for fixed
$m_{1/2}$ values $vs.$ $m_0$ in Fig. \ref{fig:stauco}.
In frame {\it a}), it is evident that the stau co-annihilation corridor
yields a relic density in accord with WMAP results for
$m_{1/2}\alt 900$ GeV. Thus, the LHC would completely explore
the stau
co-annihilation corridor for $\tan\beta =10$. 

The other region allowed by relic density constraint occurs at very
large $m_0$ where $\mu$ becomes small: the HB/FP
region\cite{hyper,fpoint}.\footnote{ The entire HB/FP region is not
shown here: over part of this region, the evaluation of the $\mu$
parameter using ISAJET 7.64 is numerically unstable. This is corrected
in ISAJET v7.65.}  Here, the growing higgsino component of the
neutralino allows for efficient annihilation into vector boson
pairs\cite{bb2,fmw}.  For very small $\mu$ values, then
neutralino-chargino co-annihilation becomes important\cite{z1w1co}. In
this FP/HB region squarks are very heavy, and as $m_{1/2}$ increases,
$m_{\tg}$ increases as well so that strongly interacting sparticle
production cross sections decrease.  Since $\mu$ is small, the light
charginos and neutralinos are significantly higgsino-like, and can only
be produced via electroweak interactions. We see that the reach of the
LHC is limited to $m_{1/2}\sim 700$ GeV.  For even higher $m_{1/2}$
values, and staying in the HB/FP region, one enters an area that is not
accessible to LHC experiments, at least via the general purpose
search strategies described here, even though
the light charginos and neutralinos are expected to have masses
less than 250-500 GeV. It would be interesting to examine whether
it is possible to devise search strategies that exploit specific
characteristics of the FP/HB region to extend the LHC reach. For
instance, in the small $|\mu|$ region, 
decays of gluinos into third generation fermions are enhanced resulting
in events with hard $b$ and $t$ quarks from the decay of heavy
gluinos. Whether it is possible to extend the LHC reach in these event
topologies merits further investigation. 

\FIGURE[t]{\epsfig{file=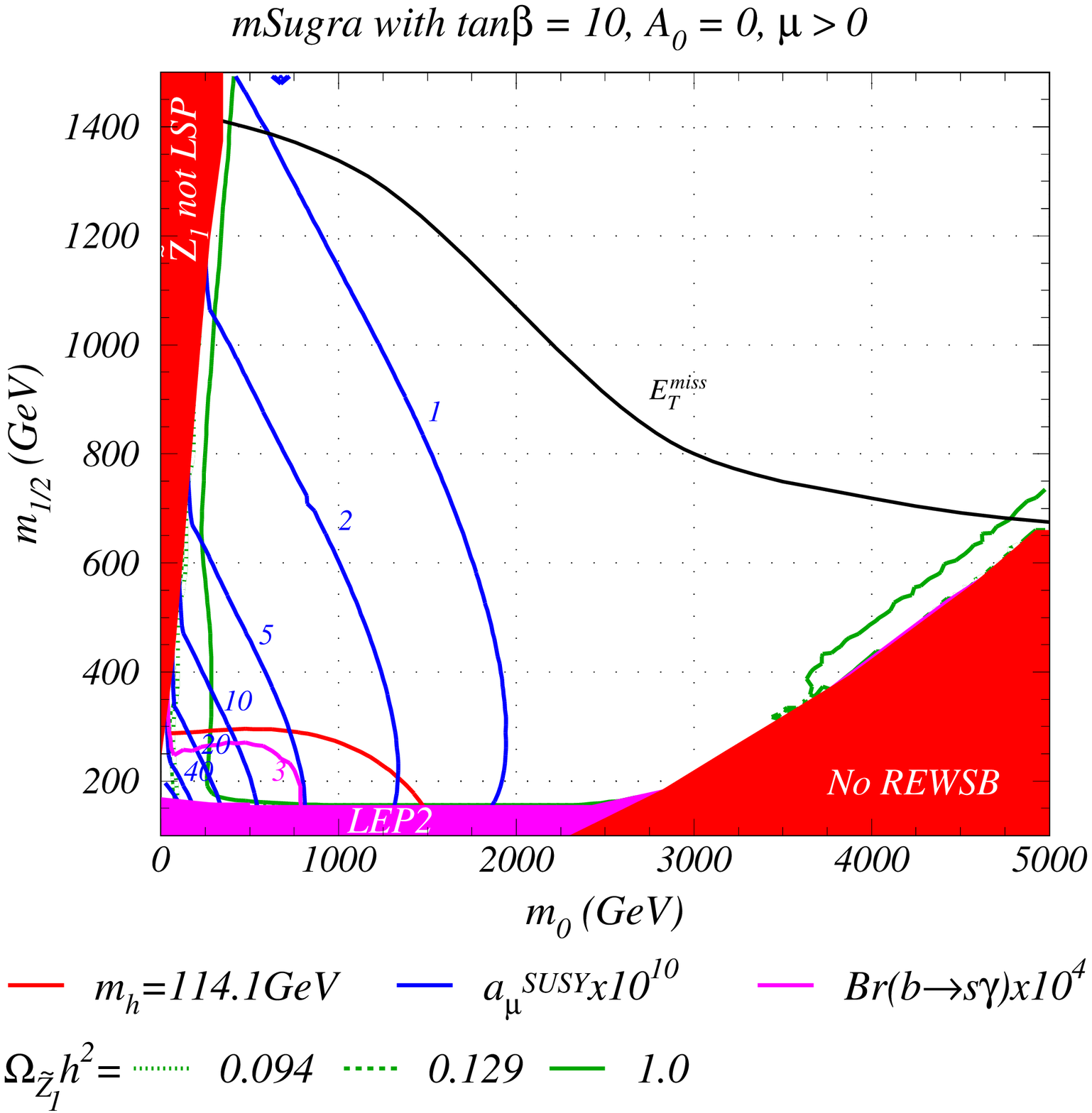,width=15cm} 
\caption{Contours of several low energy observables in the $m_0\ vs.\
m_{1/2}$ plane of the mSUGRA model, for $\tan\beta =10$, $A_0=0$ and
$\mu >0$.  We show contours 
of CDM relic density (green), together with a contour
of $m_h=114.1$ GeV (red), contours of muon anomalous magnetic moment $a_\mu$
($\times 10^{10}$) (blue) and contours of $b\to s\gamma$ branching fraction
($\times 10^{4}$) (magenta). Also shown is the maximal reach of the CERN LHC in
the $\eslt +$jets channel for 100 fb$^{-1}$ of integrated luminosity.}

\label{fig:sug10p}}

\FIGURE[t]{\epsfig{file=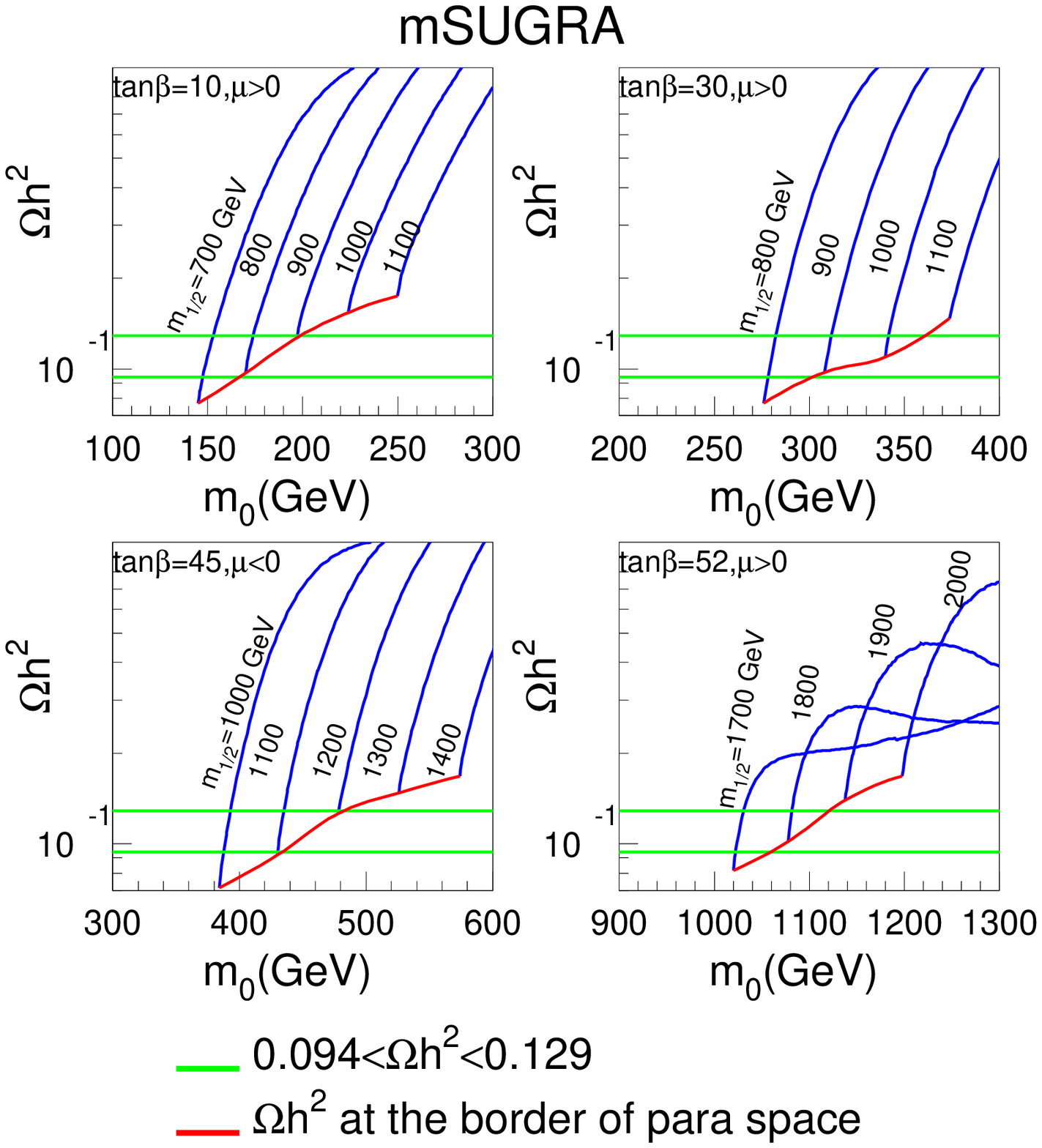,width=15cm} 
\caption{The relic density $\Omega_{\tz_1}h^2\ vs.\ m_0$ for
various fixed $m_{1/2}$ values in the stau co-annihilation
corridor, for {\it a}) $\tan\beta =10$, $\mu >0$, {\it b})
$\tan\beta =30$, $\mu >0$, {\it c}) $\tan\beta =45$, $\mu <0$
and $\tan\beta =52$, $\mu >0$. We also show via green lines the
$2\sigma$ WMAP
limits on $\Omega_{\tz_1}h^2$.}
\label{fig:stauco}}

Next, we show the contours for the same 
low energy observables as in Fig.~\ref{fig:sug10p}, but
for $\tan\beta =30$, $A_0=0$ and $\mu >0$
in Fig. \ref{fig:sug30p}. The larger value of $\tan\beta$ causes
the values $a_\mu$ and $BF(b\to s\gamma )$ to deviate more from
their SM values at low $m_0$ and $m_{1/2}$; the bulk region
of relic density annihilation is disfavored by $BF(b\to s\gamma )$. 
This leaves only the
stau  co-annihilation corridor and the HB/FP region as 
phenomenologically viable. The LHC reach contour remains similar to that 
for $\tan\beta =10$. Again, we see that if SUSY has parameters in this
plane, then LHC should see it  unless 
$m_{1/2}>700$ GeV with SUSY in the HB/FP region. In particular, the LHC
should be able to cover the stau co-annihilation corridor for $\tan\beta =30$, 
since it yields
a relic density $\Omega_{\tz_1}h^2<0.129$ for $m_{1/2}\alt 1050$ GeV,
while the LHC reach for low $m_0$ extends to $m_{1/2}\sim 1400$ GeV.

\FIGURE[t]{\epsfig{file=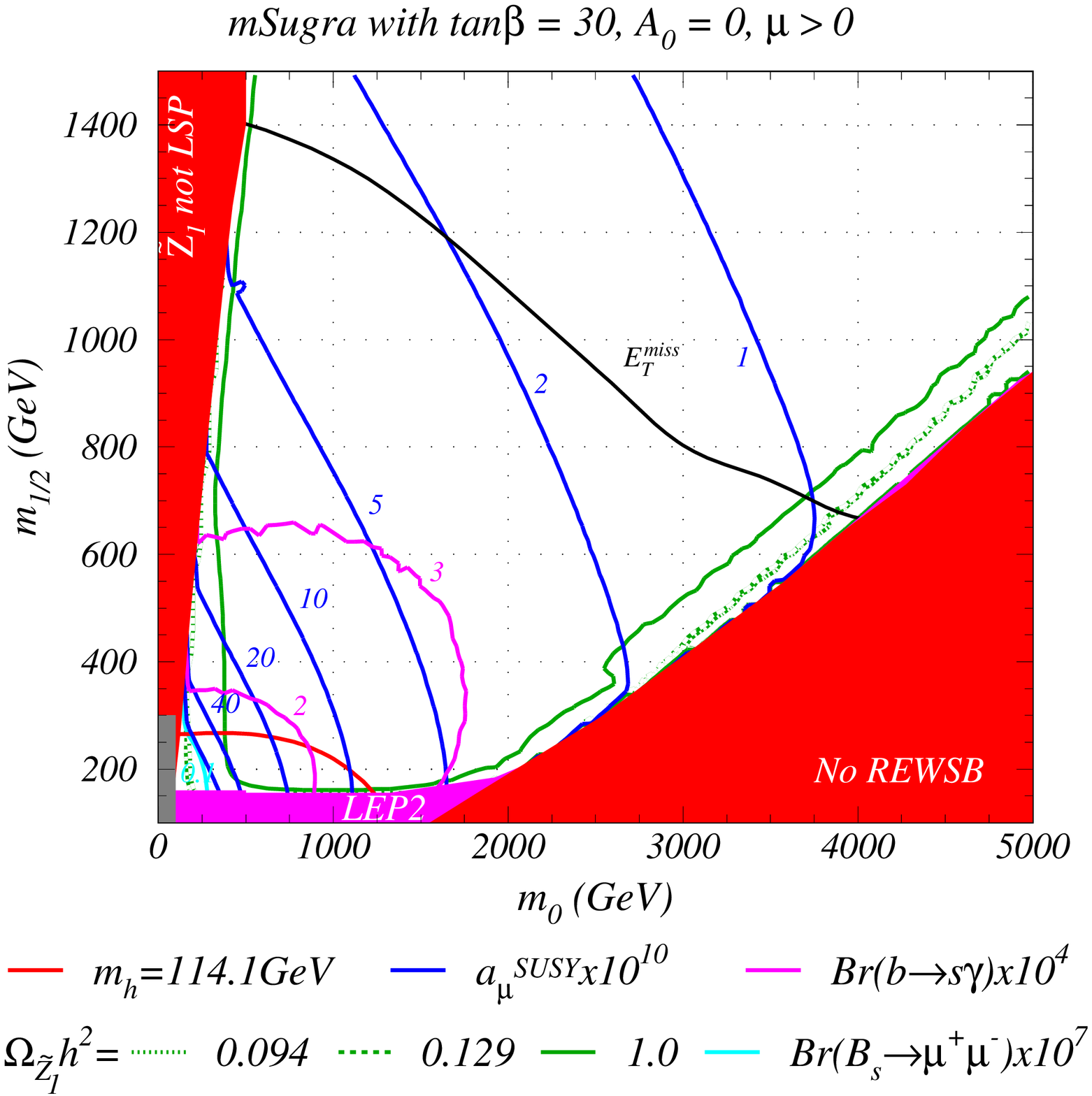,width=15cm} 
\caption{The same as Fig.~\ref{fig:sug10p} but for
$\tan\beta =30$.}

\label{fig:sug30p}}

Fig. \ref{fig:sug45m} illustrates the impact of the 
low energy observables, this time
for $\tan\beta =45$ and $\mu <0$. The LHC reach is similar
to the lower $\tan\beta$ cases, but now the low $m_0$ and $m_{1/2}$ 
bulk region is firmly excluded by both $a_\mu$ and
$BF(b\to s\gamma )$, which are large negative, and large positive, 
respectively. We see, however, that a new region of low relic density
has opened up: the $\tz_1\tz_1\to A,\ H\to f\bar{f}$
annihilation corridor, where annihilation takes place especially
through the very broad $A$ width\cite{bb2,Afunnel}. 
In fact, we see that although the
$A$ annihilation corridor extends to very large $m_{1/2}$
values, the region allowed by WMAP is almost entirely
accessible to LHC searches. A modest additional integrated luminosity
beyond 100 fb$^{-1}$ assumed in this study
should cover this entire region.
In addition, the stau co-annihilation corridor remains consistent
with WMAP constraints up to $m_{1/2}$ values as high as 1200 GeV
as shown in frame {\it c}) of Fig. \ref{fig:stauco}, 
so that LHC should be able to completely explore this region.
Finally, the HB/FP region can again be explored up to
$m_{1/2}\sim 700$ GeV at the LHC. 

\FIGURE[t]{\epsfig{file=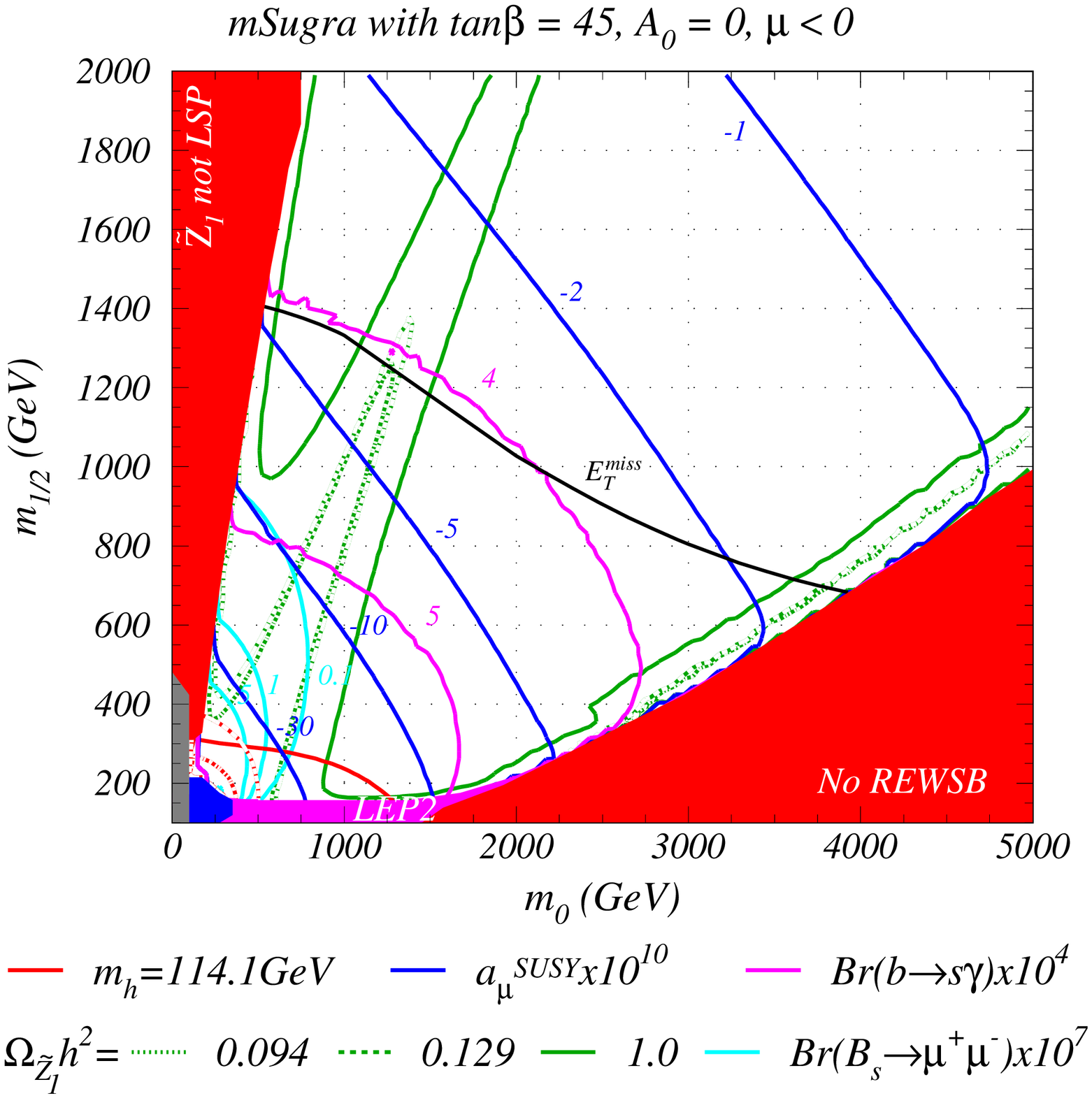,width=15cm} 
\caption{The same as Fig.~\ref{fig:sug10p} but for
$\tan\beta =45$ and $\mu<0$.}
\label{fig:sug45m}}

In Fig. \ref{fig:sug52p}, we show the same low energy contours 
in the $m_0\ vs.\ m_{1/2}$ plane, but now for
$\tan\beta =52$ and $\mu >0$. In this case, the low $m_0$ and $m_{1/2}$
bulk region is largely excluded because $BF(b\to s\gamma )<2\times 10^{-4}$.
The WMAP constraints again restrict us to either the stau
co-annihilation region at low $m_0$, or the HB/FP region at large $m_0$.
We can see from Fig. \ref{fig:stauco}{\it d} that 
the stau co-annihilation corridor extends to $m_{1/2}\sim 1870$~GeV, which
is somewhat beyond the reach of
the LHC for 100 fb$^{-1}$ of integrated luminosity. 
Models with these high parameter values suffer from
considerable fine-tuning, since the $\mu$ parameter,
which has been invoked
as a measure of fine-tuning\cite{hyper}, 
is beyond 1600 GeV for $m_{1/2}>1400$ GeV, 
so that $\mu^2/M_Z^2$ is large.
In the HB/FP 
region, the LHC reach, via these channels, again cuts off 
at $m_{1/2}\sim 700$ GeV: once again, it would be worth exploring
whether the SUSY signal can be picked up in other channels.
If we increase $\tan\beta$ beyond 52, then the $A,\ H$-annihilation
funnel re-enters the figure\cite{sug_chi2}. In Ref. \cite{sug_chi2}, 
it is shown that this annihilation funnel can reach
$m_{1/2}$ values as high as 1400 GeV for $m_0\sim 3.5$ TeV
and $\tan\beta =56$.
Thus, in this case, the LHC will not be able to access the complete 
$A,\ H$ annihilation funnel.

\FIGURE[t]{\epsfig{file=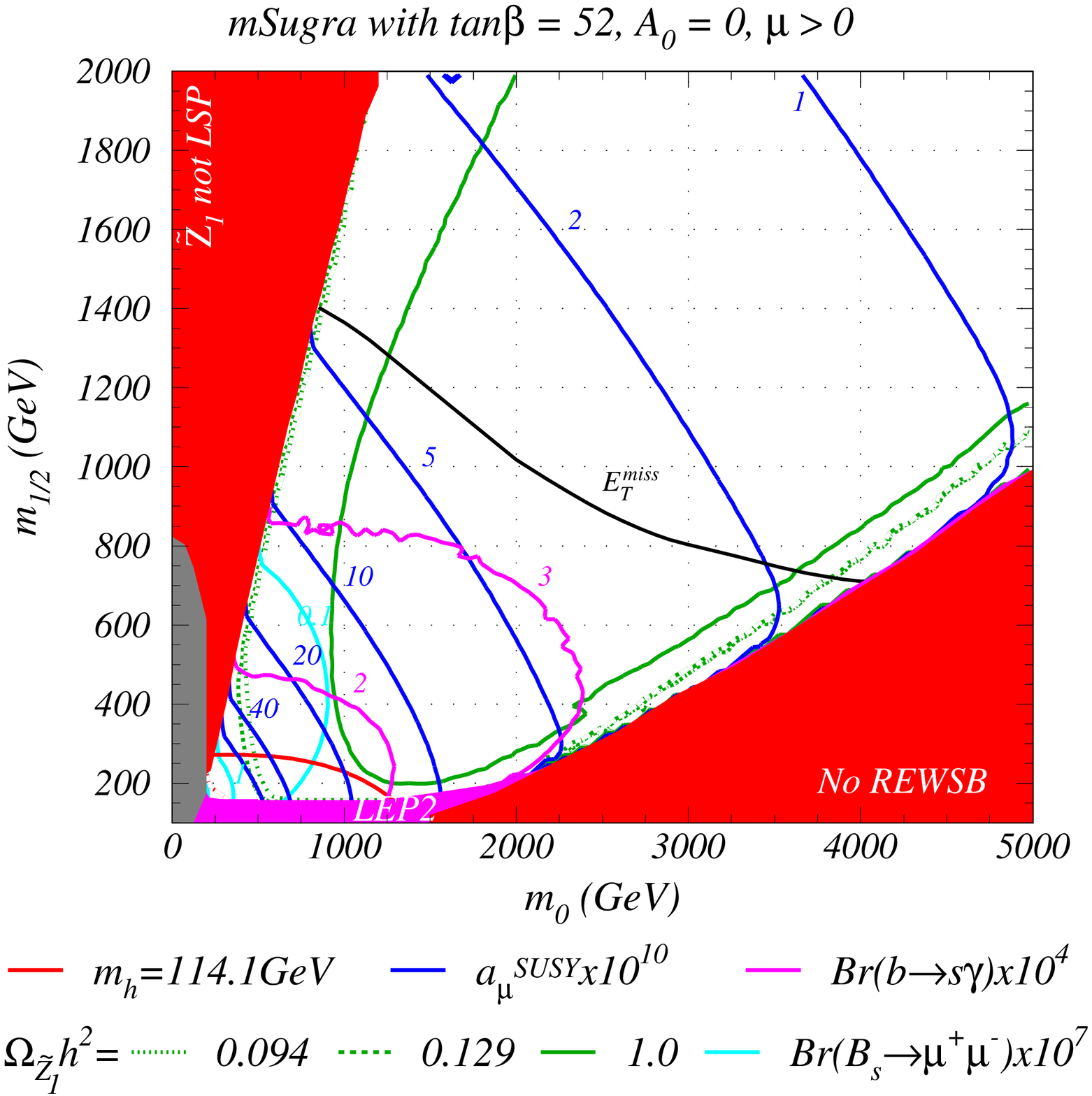,width=15cm} 
\caption{The same as Fig.~\ref{fig:sug10p} but for
$\tan\beta =45$.}
\label{fig:sug52p}}

\section{Conclusions}
\label{sec:conclude}

We have updated our assessment of the SUSY reach of the CERN LHC via
$E_T^{miss}$  and multilepton channels,
and presented new results for the reach in channels with isolated
photons or leptonically decaying $Z$ bosons. 
We work within the framework 
of the mSUGRA model, and use ISAJET v7.64 together with the CMSJET
fast detector simulation to model the CMS detector, and assume an
integrated luminosity of 100 fb$^{-1}$. 
Our results are presented over an expanded
mSUGRA model parameter space to include the reach in the
so-called HB/FP region at very large $m_0$, 
This region, together with the stau coannihilation corridor, and the
annihilation funnel where LSPs annihilate via the $A$ or $H$ resonances,
are strongly preferred by the recent data of the WMAP
collaboration. The overall LHC reach turns out to be
quite insensitive to $\tan\beta$. We find that experiments at the LHC
will probe
$m_{1/2}\alt 1400$ GeV for low $m_0$, and  
$m_{1/2}\sim 700$ for large $m_0$ in the HB/FP region.
These values correspond to $m_{\tg}\sim 3000$ GeV and
$1800$ GeV, respectively. 

We have also presented the reach in
a variety of multi-lepton channels. The reach in these
individual channels is, in general,
sensitive to $\tan\beta$. We also show the reach in a channel including 
reconstructed $Z^0\to\ell\bar{\ell}$ decays, and channels including
isolated photons. The isolated photon signals may contain
$h\to\gamma\gamma$ events at a low, but observable rate. Indeed the SUSY
event sample may contain SM background-free $h \to
\gamma\gamma$ events, though a very  high integrated luminosity will be
needed to identify the rate-limited $h$ signal.
In the HB/FP region, the
photonic channels also have a slight enhancement from 
radiative neutralino decays $\tz_2\to\tz_1\gamma$.
For $m_{1/2} \alt 800 (400)$~GeV and small (large) values of $m_0$,
there should be observable signals in all these channels if new physics
discovered at the LHC is to be interpreted as supersymmetry as realized in
the mSUGRA  model.

We have also examined the reach of the LHC in light of the recent assessment
of direct (from LEP2) and indirect constraints on the mSUGRA model. 
The indirect constraints include 
the neutralino relic density $\Omega_{\tz_1}h^2$ from recent WMAP
analyses together with accelerator
measurements of $BF(b\to s\gamma )$, $a_\mu =(g-2)_\mu$ and
the bound on $BF(B_s\to\mu^+\mu^- )$ (this bound is hardly constraining 
for the parameter planes that we have examined). For large values of
$\tan\beta$, experimental values of
$BF(b\to s\gamma )$ and $a_\mu$  disfavor
negative values of the $\mu$ parameter unless $m_0$ and $m_{1/2}$
are also large, but for $\tan\beta \sim 10$, values of $m_0$ and
$m_{1/2}$ $\agt 400-500$~GeV are perfectly acceptable.
For $\mu >0$, $m_0$ would have to be rather small
so that the relic density is either in the bulk annihilation
region or in the stau coannihilation strip, or $m_0$ would
have to be very large, in the HB/FP region. The CERN LHC can definitely 
explore all the bulk annihilation region, and can explore all the stau
co-annihilation corridor unless $\tan\beta$ is very high.
The LHC can explore the HB/FP region up to $m_{1/2}\sim 700$ GeV via
the conventional SUSY search channels.
However, the HB/FP region appears to extend indefinitely to large
$m_{1/2}$ and $m_0$ values, ultimately well beyond the LHC reach.
For large $\tan\beta$, the $A,\ H$ annihilation funnel enters the
$m_0\ vs.\ m_{1/2}$ plane. The LHC with 100 fb$^{-1}$ ought to be able
to detect SUSY over this entire region, except for the case of 
very large $\tan\beta\sim 56$ with $\mu >0$.

\section*{Acknowledgments}
 
We thank M. Drees for comments on the manuscript.
This research was supported in part by the U.S. Department of Energy
under contracts number DE-FG02-97ER41022 and DE-FG03-94ER40833.
	
%

\end{document}